\documentclass[journal=nalefd,manuscript=letter]{achemso}

\usepackage{epsf}
\usepackage[dvips]{color}
\usepackage{graphicx}

\author{Giovanni Cantele}
\email{Giovanni.Cantele@na.infn.it}
\affiliation{Coherentia CNR-INFM and Universit\`a di
Napoli ``Federico II'' ,
Dipartimento di Scienze Fisiche, Complesso Universitario
Monte Sant'Angelo, Via Cintia, I-80126 Napoli, Italy}
\author{Young-Su Lee}
\affiliation{Materials Science and Technology Research Division,
Korea Institute of Science and Technology, Seoul 136-791, Republic of Korea}
\author{Domenico Ninno}
\affiliation{Coherentia CNR-INFM and Universit\`a di
Napoli ``Federico II'' ,
Dipartimento di Scienze Fisiche, Complesso Universitario
Monte Sant'Angelo, Via Cintia, I-80126 Napoli, Italy}
\author{Nicola Marzari}
\affiliation{Department of Materials Science and Engineering,
and Institute for Soldier Nanotechnologies, Massachusetts
Institute of Technology, Cambridge, Massachusetts 02139, USA}

\title[Spin channels in functionalized graphene nanoribbons]
{Spin channels in functionalized graphene nanoribbons}

\begin{document}

\begin{abstract}
We characterize the transport properties of functionalized graphene nanoribbons
using extensive first-principles calculations based on density functional theory (DFT) that encompass both monovalent and
divalent ligands, hydrogenated defects and vacancies. 
We find that the edge metallic states are preserved under a variety of 
chemical environments, while bulk conducting channels can be easily destroyed
by either hydrogenation or ion or electron beams, resulting in devices that 
can exhibit spin conductance polarization close to unity.
\end{abstract}

Graphene is currently the subject of intense experimental and
theoretical investigations, thanks to remarkable and
novel physical
features~\cite{rev:CastroNetoRMP2009,rev:geim,exp:chen,exp:KimPRL2007,exp:KimAPL2007,rev:avouris}.
Linear dispersion around the Fermi energy ($E_{F}$), high crystallinity
and mobility, ballistic transport on the sub-micrometre scale (even at room temperature),
and a two-dimensional structure that is amenable to lithographic
techniques~\cite{exp:chen,exp:KimPRL2007,exp:BergerScience2006} offer great promise for
nanoelectronics applications, from field effect transistors to
interconnects~\cite{exp:Lin} to spintronics and related
applications~\cite{exp:TombrosNature2007,exp:ChoAPL2007}.

Electronic-structure
calculations have shown that (real-space) edge states
emerge in zigzag graphene nanoribbons (ZGNR)~\cite{th:nakada};
the resulting peak in the density of states
(DOS) at $E_{F}$ induces a magnetic instability that leads to a ground
state with antiferromagnetic (AFM) ordering,
where the spins on the two ribbon
edges~\cite{th:pisani,th:SonNature2006,th:SonPRL2006} have opposite orientations.
Under an applied 
electric field either half-metallic~\cite{th:SonNature2006,th:SonPRL2006} or
half-semiconducting~\cite{th:rudberg} behavior can be
observed, of extreme interest for spintronics applications~\cite{th:wang}.
However, the magnitude of the field
perpendicular to the ribbon required to close the band gap for one of 
the spins can be as large as 0.1 V {\AA}$^{-1}$,
although it is a decreasing function of the
ribbon width.
In contrast with electronic-structure
predictions~\cite{th:scuseria1}, recent measurements~\cite{exp:KimPRL2007} 
show a weak dependence of the energy gap on the crystallographic direction, 
highlighting the critical role that edge structures~\cite{th:querlioz},
passivations~\cite{th:ferrari,th:PeetersPRB2007,th:PeetersPRB2008} and 
scattering taking place at rough boundaries~\cite{th:tomanek,th:areshkin} can play.

Since edge terminations, passivations and defects play a key role
in the performance of graphene devices~\cite{th:scuseria2,th:ferrari},
we characterize in this Letter the avenues available to control electronic
and spin transport 
with chemical functionalizations or ion- or electron-beam treatments.
We study single- and double-bonded moieties (H, OH, F, Cl, Br, S, O, NO$_{2}$, and
NH$_{2}$) and hydrogenated defects and vacancies with extensive DFT
first-principles 
calculations~\bibnote{
We use density functional theory (DFT) within the Perdew-Burke-Ernzerhof approximation~\cite{th:pbe}
and the plane-wave {\sc Quantum-ESPRESSO} package~\cite{code:QE}, with a 40 Ry
cutoff for the wave functions, 480 Ry for the charge density, $16\times 1\times 1$
Monkhorst-Pack~\cite{th:mp} sampling of the BZ, electronic occupations of metallic systems
using Marzari-Vanderbilt cold smearing~\cite{th:mv} of 0.007 Ry, and ultrasoft
pseudopotentials to represent ionic cores~\cite{th:rrkj,th:van}.  13 {\AA} and 7 {\AA}
separate the ribbon periodic replica in the directions perpendicular and parallel to the
ribbon plane.  Defected ribbons and edges are built by replicating 8 times the bare ribbon
unit cell along the ribbon axis direction and the corresponding BZ sampled with a
$2\times 1\times 1$ MP grid.  The geometries of all the considered ribbons are optimized by
relaxing the atomic positions until all components of all forces are smaller than
10$^{-3}$ Ry/au.  Tests on sample
systems revealed almost full convergence of the calculated properties with respect to the
parameters used in the calculation.  Formation energies are calculated as
$\Delta E = E_{tot} - \sum_i n_i \mu_i$, where $E_{tot}$ is the total energy of the given
system and $n_i$ the number of atoms of species $i$ (with chemical potential $\mu_i$) in
the unit cell. We used as reference the total energy per atom of bulk graphene, H$_2$,
spin polarized O$_2$, F$_2$, and the S$_8$ molecule.

It should be pointed out that DFT has been proven to correctly describe the
properties of graphene and carbon-based nanostructures in the presence of
different chemical
environments\cite{th:PeetersPRB2008,th-exp:LiPRL2006,th-exp:JinPRL2009,th-exp:WehlingNanoLett2008},
with LDA and GGA results usually bracketing more accurate estimates obtained
with hybrid functionals. While a proper description of optical properties often
requires the introduction of many-body corrections\cite{th:prezzi,th:darancet},
quantitative agreement with conductance measurements has been obtained even
within a tight-binding framework\cite{exp:Lin}.}.
We use an original approach suitable to describe the electronic structure and
ballistic transport in large-scale nanostructures~\cite{th:calzo,th:lee},
based on the chemically accurate and minimal basis of maximally-localized Wannier
functions (MLWFs)~\cite{th:marzari,th:souza,th:mostofi}.

\begin{figure}[t!]
\epsfxsize=8.0cm \epsfbox{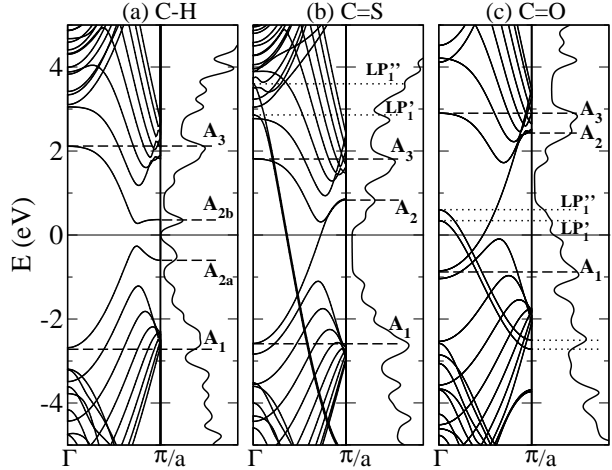}
\caption{\label{fig:bands} Band structure 
and DOS for
the AFM ground state of
(a) H:ZGNR-8, (b) S:ZGNR-8 and (c) O:ZGNR-8.
Dashed and dotted lines and labels identify some relevant band edges (see text).
The energy scale is such that E$_F$=0 eV.}
\end{figure}

First, we consider in \ref{fig:bands} the band structure and DOS of the ground state
for the representative single- and double-bonded ribbons H:ZGNR-8, S:ZGNR-8 and O:ZGNR-8 (we use here
the convention of indicating the species functionalizing the edge of the GNR first,
and the width of the ribbon in graphene units last).
As widely reported, 
the ground state for H:ZGNR-$N$ is AFM and insulating, with the two bands close to $E_{F}$
originating from bonding and antibonding Bloch sums of the $p_z$
carbon orbitals.
A similar description applies to other monovalent passivating species, such as OH or F, or
even to more complex electropositive or electronegative ligands, such as
NH$_2$ and NO$_2$, since in these latter cases the bands originating from the lone
pairs remain below the top of the valence.
We label the band edges generated by the $p_z$ orbitals
$A_1$, $A_{2a,b}$ and $A_3$ (see \ref{fig:bands}).
As $k$ goes from the Brillouin-zone (BZ) center to the edge, 
the states in these two bands evolve from being delocalized in 
real space inside the 
ribbon (i.e. in the ``bulk'') to becoming localized at the two ribbon edges.
As it will be shown later, such observation is critical to 
engineer the balance
between unpolarized bulk ribbon conductance and the conductance channels at
the ribbon edges.

Double-bonded ligands, such as O and S (\ref{fig:bands}(b) and (c)), give rise 
to a different picture:
the orbitals of the passivating species undergo 
$sp^2$ hybridization, with
one $sp^2$ orbital contributing to a $\sigma$
carbon-ligand bond and the other two 
accommodating two lone pairs; the remaining $p_z$ orbital 
participates in the $\pi$ carbon-ligand bond.
Remarkably, the semiconducting character of the hydrogenated ZGNR is reverted to
metallic in presence of these edge passivations.
Indeed, for these double-bonded functionalizations 
the $A_1-A_2$ and $A_3-A_2$ bands merge even in the AFM
ground state, but at energies higher than $E_{F}$; this can be contrasted with non-magnetic (NM) H-ZGNRs,
where the bands merge exactly at $E_{F}$.
The fully occupied $A_1-A_{2a}$ band of AFM H:ZGNR-8 becomes a mostly occupied band for
S passivations (\ref{fig:bands}(b)) and a mostly empty band for O passivations
(\ref{fig:bands}(c)) - the more electronegative
oxygen is extracting charge from the ribbon.
The double-bonded moieties show two new bands
(whose extrema at $k=0$ are denoted by $LP_1^{\prime}$ and $LP_1^{\prime\prime}$)
that remain very close in energy across most of the BZ.
Direct inspection of their charge density shows a marked lone-pair 
character, and so they are the analogue of the lone-pair 
bands alluded to before for other monovalent functionalizations (e.g. NH$_2$), with the
key difference that these now
cross the Fermi level.
Differences between the AFM ground state and the NM solution are much less pronounced for
these double-bonded moieties than for H:ZGNR-$N$
(see also later, \ref{tab:form_en}),
the only effect being a small splitting arising for 
these two lone-pair bands.
Indeed, the origin of the magnetic instability, arising in H:ZGNR-$N$ from the large DOS of
the NM state at $E_{F}$, disappears in the presence of double-bonded species
(\ref{fig:bands}(b), (c)).
Charge density plots for the A$_1$-A$_2$-A$_3$ bands reveal, for O:ZGNRs, an opposite behavior 
with respect to hydrogenated ribbons:
eigenstates near $k=0$ are mostly localized onto edge O atoms,
whereas localization in the bulk of the ribbons is found towards the zone boundary.

\begin{figure}[t!]
\epsfxsize=8.0cm \epsfbox{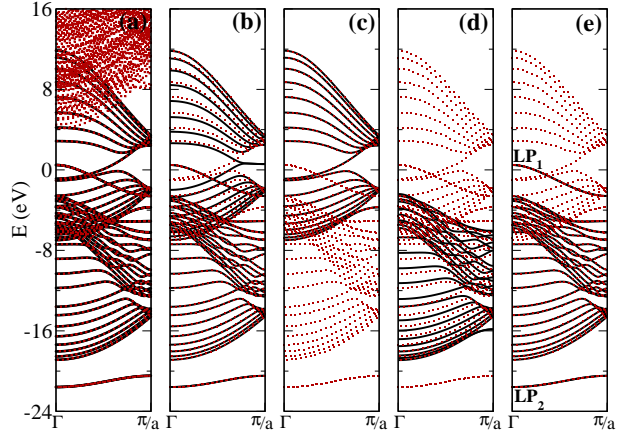}
\caption{\label{fig:WFs}  Band structure 
(red dots) and its MLWFs interpolation (solid black lines) for 
O:ZGNR-8, obtained by including, respectively: 
(a) all MLWFs (see text), (b) all MLWFs but the $p_z$ MLWFs
on the oxygens, (c) all $p_z$ MLWFs, (d) all $\sigma$ MLWFs, and
(e) all $\sigma$ MLWFs and the lone-pair $sp^2$ MLFWs. NM state
is chosen for clarity.}
\end{figure}

Clear chemical insight on the nature of all the energy bands is obtained after we extract
MLWFs~\cite{th:lee,th:marzari,th:souza,code:w90} from the occupied and the $\pi$ unoccupied manifolds of these ribbons; we choose here as an example
the NM state of O:ZGNR-8. The resulting MLWFs map the Bloch bands onto an explicit 
``tight-binding'' basis of localized orbitals, composed of one $p_z$ MLWF for every C and for every O,
one $\sigma$ bonding MLWF for every C-C and for every C=O bond, and two lone-pair $sp^2$ MLWFs for every O. 
The band structure of the ribbon in the full
BZ can then be straightforwardly interpolated across the BZ by diagonalizing the Hamiltonian in this minimal
basis set~\cite{th:lee,th:yates}, or in different subsets.
We show in \ref{fig:WFs} our results: the first panel compares the band structure obtained by diagonalization in
the minimal MLWFs basis and the full calculation in a complete plane-wave basis set, showing that the manifolds
of interest are all reproduced with extreme accuracy.
The central role played by the oxygen $p_z$ MLWFs is highlighted
in panel (b):
removing these from the minimal basis 
is sufficient to destroy the agreement around $E_{F}$, resulting instead in energy bands similar to those
of NM H:ZGNR-8, where the $A_1-A_2$ and $A_3-A_2$ bands merge, in proximity of the BZ boundary, at
$E_{F}$.
The manifold obtained by all the $p_z$ (including those on the oxygens) is shown in panel (c), while the
$\sigma$ manifold is in panel (d); in this latter case, quantitative agreement is obtained only after
the lone-pair $sp^2$ MLWFs on the oxygens are added to the $\sigma$ basis set (panel (e)), since these lone pairs
not only give rise to the two lone-pair bands $LP_1$ and $LP_2$, but are needed to bring the $\sigma$ manifold in
full agreement with the exact reference result. The
two manifolds in (c) and (e) represent a virtually exact 
decomposition of the original band structure, highlighting once again the role
of MLWFs in capturing a faithful physical picture in the smallest possible
representation.

\begin{table}
\begin{tabular}{c|c|c|c|c}
\hline\hline
ribbon    & $E_f^{AFM}$ & $E_{NM}-E_{AFM}$ & $E_{FM}-E_{AFM}$ & $\mu_{FM}$ \\
\hline
OH:ZGNR-4 & -3.38       & 0.036            & 0.011            & 0.37       \\
 F:ZGNR-4 & -2.70       & 0.047            & 0.014            & 0.35       \\
 H:ZGNR-4 &  0.38       & 0.059            & 0.014            & 0.39       \\
 H:ZGNR-6 &  0.42       & 0.081            & 0.015            & 0.45       \\
 H:ZGNR-8 &  0.42       & 0.083            & 0.007            & 0.62       \\
 O:ZGNR-8 & -1.03       & 0.008            & 0.000            & 0.19       \\
 S:ZGNR-8 &  0.99       & 0.002            & 0.000            & 0.17       \\
\hline\hline
\end{tabular}
\caption{\label{tab:form_en} Unit-cell ground-state formation energy ($E_f^{AFM}$), magnetic instability
($E_{NM}-E_{AFM}$), magnetic interaction strength ($E_{FM}-E_{AFM}$) and magnetic
moments in the FM state ($\mu_{FM}$) for ribbons of different widths and passivations. 
Energies are in eV, magnetic moments in Bohr magnetons.}
\end{table}

\begin{figure*}[b!]
\epsfxsize=7.8cm\epsfbox{DOS_defect.eps}\hspace*{0.2cm}
\epsfxsize=1.25cm\epsfbox{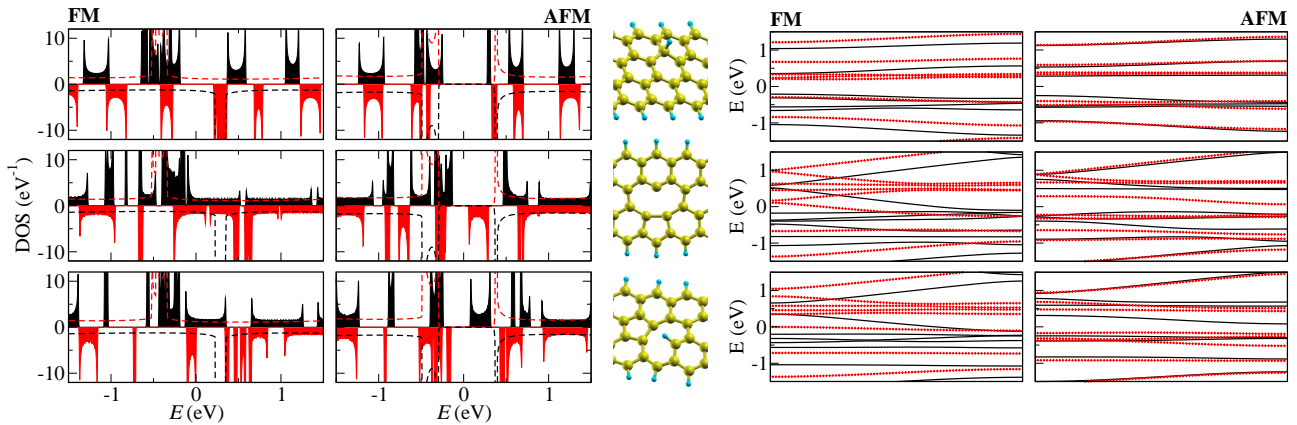}\hspace*{0.2cm}
\epsfxsize=7.45cm\epsfbox{defect_band_comparison_paper.eps}
\caption{\label{fig:DOS} (left) DOS of a H:ZGNR-4 (both FM
and AFM states) in the presence
of several periodic defects (one defect for 8 ZGNR unit cells):
on-top H (top panels), C vacancy (central panels), and  
hydrogenated C vacancy (bottom panels). The local structure of the defects in shown
in the middle inset. Dashed lines correspond to the pristine ribbon, ``negative'' DOS
is for the minority spin.
(right) Band structures for both FM and AFM states, black solid lines (red diamonds)
 correspond to the majority (minority) spin.}
\end{figure*}

The two lone-pair bands, originating from the $sp^2$ MLWFs, are very 
dispersive, with a bandwidth of
$~$3 eV for O:ZGNR-8 and $~$9 eV for S:ZGNR-8;
the larger size of the sulphur is responsible for the 
broader dispersion, 
with the largest hopping element in the Hamiltonian being
-0.9 eV for O:ZGNR-4, and -2.94 eV for S:ZGNR-4. 
Nevertheless, there is no instability toward dimerization at 
the edges, even if
the bond lengths for O-O and S-S of 1.48 and 2.05 {\AA}
are comparable to the 2.46 {\AA} spacing along the ribbon edge.
Only halogenation with larger species (Cl and Br) 
gives rise to unit-cell doubling, 
with neighboring halogens assuming opposite tilts 
with respect to the ribbon plane.

The thermodynamic stability of the different ribbons is summarized in 
\ref{tab:form_en}: negative formation energies are found for OH, F and O passivations.
The AFM state is the ground state for all the systems considered, but the energy difference between
the FM and AFM states decreases with the ribbon width (see \ref{tab:form_en} and
Ref.~\cite{th:pisani}). 
Room-temperature FM coupling between the edges could be stabilized by an applied magnetic field or
induced by defects or adsorbates~\cite{th:pisani}; adsorbates can also break the spin-up/spin-down 
symmetry in the AFM state, inducing
half-semiconductivity~\cite{exp:defects1,exp:defects2,exp:defects3,exp:defects4,exp:defects5}.
Spin stiffness along the edges, as reported recently~\cite{th:yazyev},
is remarkably high, raising hopes of room-temperature coherence for these
one-dimensional systems at length-scales comparable to those of microelectronics devices.

\begin{figure}[t!]
\epsfxsize=8.0cm \epsfbox{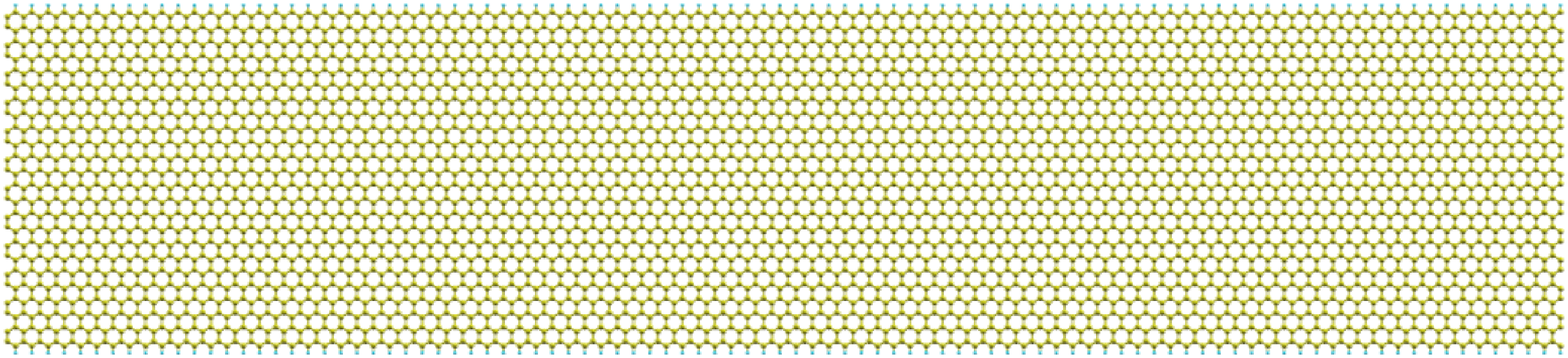}\vspace*{0.2cm}
\epsfxsize=8.0cm \epsfbox{ZGNR-24_cond_random_px_removed_FM.eps}\hspace*{0.2cm}\epsfxsize=8.0cm \epsfbox{ZGNR-24_cond_random_px_removed_AFM.eps}
\caption{\label{fig:cond24}  Top panel:
structural model of the long ZGNR, used for conductance calculations.
Left bottom panel:
spin-up ($G_\uparrow$) and spin-down ($G_\downarrow$) conductance and
conductance polarization ($P_G$) of an infinite H:ZGNR-24 in the FM
state, whose central section ($\sim$197 {\AA}) has been randomly hydrogenated. 
Averages (solid thick lines) are taken over 30
random configurations (solid thin lines).
Shaded (grey) regions represent, at each energy, the standard deviation.
Dashed (red) curves represent the same quantities for the pristine
ribbon. Right bottom panel: same as in (a) for the AFM state.}
\end{figure}
A breakthrough application for graphene ribbons would be in the role of spin valves; indeed, 
energy windows where electronic states with only one spin are available occur under 
different circumstances. We first explore this point in \ref{fig:DOS},
where we compare the DOS and band structures of three 
defected H:ZGNR-4 in the presence of bulk ribbon hydrogenations, 
carbon vacancies (as e.g. induced
by ion irradiation) or both; in all cases the breaking (enhancement)
of the spin-up/spin-down symmetry (asymmetry) of the AFM (FM) state is observed.
On the other hand, in any real device no degree of order 
along the ribbon can be expected, and
spin-polarized currents along the edges would be overshadowed 
by the spin-unpolarized transport channels available inside the 
bulk of the ribbon; around $E_{F}$, these are dominated by the
MLWFs $p_z$ contributions detailed before.
Thus, engineering a ZGNR device for spintronics applications 
will require removing the unpolarized bulk conduction channels while 
preserving edge conductance. This could be achieved with hydrogenations,
or ion or electron beams;
these are all now recognized 
as effective tools to tailor electronic and 
transport properties~\cite{exp:defects1,exp:defects2,exp:defects3,exp:defects4,exp:defects5}, and
proton irradiation even induces ferromagnetism in 
NM graphitic samples~\cite{exp:defects1,exp:defects2,exp:defects3,exp:defects4,exp:defects5}.

\begin{figure}[t!]
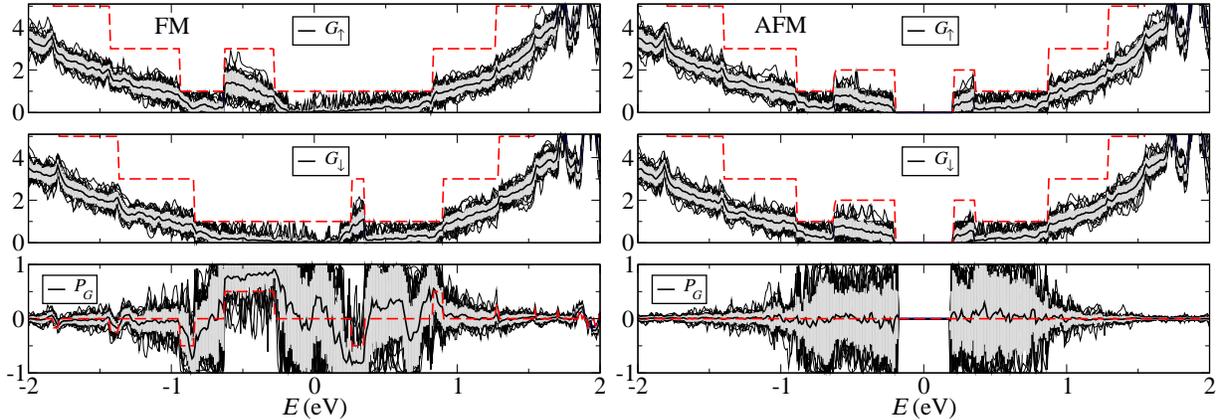

\epsfxsize=8.0cm \epsfbox{ZGNR-12_cond_random_px_removed_FM.eps}\vspace*{0.2cm}
\epsfxsize=8.0cm \epsfbox{ZGNR-12_cond_random_px_removed_AFM.eps}
\caption{\label{fig:cond12} 
Top panel:
spin-up ($G_\uparrow$) and spin-down ($G_\downarrow$) conductance and
conductance polarization ($P_G$) of an infinite H:ZGNR-12 in the FM
state, whose central section ($\sim$197 {\AA}) has been randomly hydrogenated. 
Averages (solid thick lines) are taken over 30
random configurations (solid thin lines).
Shaded (grey) regions represent, at each energy, the standard deviation.
Dashed (red) curves represent the same quantities for the pristine
ribbon. Bottom panel: same as in (a) for the AFM state.}
\end{figure}

We investigate here this possibility, and show that
the availability of spin-polarized edge states allows for the design of
simple and robust spin-valves. Our strategy is based 
on the observation that most or all functionalizations 
to the bulk of the ribbon induce $sp^2$ to $sp^3$ re-hybridization,
and remove the ``half-filled'' $p_z$ MLWFs from the energy window
around $E_F$ \cite{th:lee,th:lee2}.
This mechanism is central to the realization of a spin valve where
unpolarized conduction channels are removed from the bulk of the ribbon, while
preserving at the same time the edge states.
Chemical reactions with atomic hydrogen represent the simplest 
route, and we show in \ref{fig:cond24} the quantum conductance calculated 
for a realistic device: an
infinite H:ZGNR-24 ($\sim$50 {\AA} width) whose central section (3840 atoms,
$\sim$197 {\AA}) has 
been randomly hydrogenated with a 0.5\% defect concentration.
The conductances $G_\uparrow$ and 
$G_\downarrow$ for both spin channels
as well as the conductance polarization $P_G$ (defined as
$P_G=(G_\uparrow-G_\downarrow)/(G_\uparrow+G_\downarrow)$) are averaged over
30 configurations, each with random hydrogenations.
For comparison, we show in \ref{fig:cond12} the same quantities
for the H:ZGNR-12 ribbon (the central section now contains 1920 atoms).
In the FM state, 
the spin valve character of such system is immediately apparent, with
a pronounced spin up or spin down conductance peak just below or above
$E_{F}$, where no significant contribution from the other spin is found.
Therefore, tuning the applied bias voltage will turn on and
off the injection of electrons with a given spin with an efficiency that
may reach 100\% for single devices. 
In the AFM state,
while a random distribution
of defects would induce a spin polarization due to the breaking of
the symmetry between spin up and spin down channels,
the average effect is zero. Yet, mesoscopic fluctuations 
mean that for each and every device
spin valve effects would be very relevant even in this case, as 
already inferred for graphene
ribbons with rough edges~\cite{th:tomanek}. Such effects appear to be
more relevant for the smaller ribbon (\ref{fig:cond12}).
Irradiation with ion or electron beams,
that nowadays can be focused onto areas as small as 
a few nm or a few {\AA} in diameter respectively, 
would lead to similar effects, since the creation of vacancies 
(hydrogenated or not) in the bulk of the ribbon also destroys the 
unpolarized conduction channels (see \ref{fig:DOS}). Provided one edge of the ribbon can be
physically protected, a wide variety of chemical functionalizations - starting
from double hydrogenations at the opposite edge - would achieve the same
objective.

In conclusion, we have calculated the 
electronic-structure and quantum conductance
in realistic, functionalized graphene nanoribbons with full 
chemical accuracy, thanks to the use of
an accurate but minimal MLWFs basis. Defects 
in spin-polarized ZGNRs - following hydrogenation or 
treatment with ion or electron beams -
can be effectively used to remove the unpolarized conduction channels in the
bulk of the ribbon, while preserving
edge states. Similar results can be devised using chemical routes
for graphene functionalization/hydrogenation~\cite{exp:ryu}. For instance,
very recently the reversible hydrogenation
of a graphene sheet has been experimentally proven, leading to the
$sp^2$ (graphene) to $sp^3$ (graphane) transformation of the carbon
network~\cite{exp:elias}.
The resulting asymmetry between the spin-up and spin-down
channels makes these systems ideal candidates for spin-polarized 
transport with a very high degree of spin polarization.  

\begin{acknowledgement}
Financial support from 
CNR ``Short mobility program 2006'', MIUR-PRIN-2007, NSF DMR-0304019, and the 
IFC Focus Center, and
computational resources from CINECA (``Progetti Supercalcolo 2008'')
are gratefully acknowledged.  
\end{acknowledgement}

\bibliography{GNRs}

\end{document}